\documentclass[seceq]{ptptex}

\usepackage{graphicx}

\newcommand{\bm}[1]{\mbox{\boldmath$#1$}}

\notypesetlogo                       
\title{
Fermionic Collective Modes in QGP near Critical Temperatures%
}

\author{
Yukio \textsc{Nemoto}$^{1,}$\footnote{ e-mail address:
nemoto@hken.phys.nagoya-u.ac.jp}  
,
Masakiyo \textsc{Kitazawa}$^{2,}$\footnote{ e-mail address:
kitazawa@quark.phy.bnl.gov} 
,
Tomoi \textsc{Koide}$^{3,}$\footnote{ e-mail address:
koide@if.ufrj.br} 
and
Teiji \textsc{Kunihiro}$^{4,}$\footnote{ e-mail address:
kunihiro@yukawa.kyoto-u.ac.jp} 
}

\inst{
$^{1}$ Department of Physics, Nagoya University, Nagoya 464-8602,
Japan\\
$^{2}$ RIKEN-BNL Reseach Center,
Brookhaven National Laboratory, Upton, NY 11973, USA\\
$^{3}$ Instituto de F\'isica, Universidade Federal do Rio de Janeiro,
C. P. 68528, 21945-970, Rio de Janeiro, Brasil\\
$^4$ Yukawa Institute for Theoretical Physics,
Kyoto University, Kyoto 606--8502, Japan\\
}

\abst{
We investigate the quark spectrum in the quark-gluon plasma phase near
color superconducting (CS) and chiral phase transitions.
Owing to the precursory soft modes of the phase transitions, there appear
novel excitaion spectra:
In the CS transition, the quark matter shows non-Fermi liquid behavior and
leads to the pseudogap in the density of states of quarks.
In the chiral transition, three collective excitations appear in the quark
spectrum.
}

\begin{document}

\maketitle

\section{Introduction}

Recently, the quark-gluon plasma (QGP) just above the chiral and
deconfinement phase transition is believed to be an unexpectedly
strongly interacting system, which is based on 
the facts that the created matter at RHIC behaves like a perfect 
fluid and that some hadronic bound states of heavy quarks can survive
above the critical temperature ($T_c$) from Lattice QCD.
Because the fundamental degrees of freedom in QGP are quarks and gluons,
it is also important to study their properties in such a strongly interacting
system.
Here, we investigate the quark spectrum just above $T_c$ of 
the chiral transition
at zero density, and the color superconducting (CS) transition around 
$\mu_B=1$ GeV with $\mu_B$ being the baryon number density,
focusing on the precursory soft modes of these transitions.
It is known that these soft modes exist over a wide range of
temperature above $T_c$ owing to a strong coupling nature between 
quarks.\cite{Hatsuda:1985eb,Kitazawa:2001ft}
In this paper, we show that they affect the quark spectrum significantly 
in a region
just above $T_c$.

\section{Precursory soft modes} \label{soft}

To describe the quark matter near $T_c$, we employ the 
two-flavor Nambu--Jona-Lasinio type interaction with the scalar
diquark correlation included,
$
{\cal L} = \bar{\psi} i/\hspace{-0.2cm}\partial  \psi
+ G_S [(\bar{\psi}\psi)^2
+ (\bar{\psi} i\gamma_5 \vec{\tau}\psi)^2]$
$ 
+G_C
(\bar{\psi} i\gamma_5 \tau_2 \lambda_A \psi^C)
(\bar{\psi}^C i\gamma_5 \tau_2 \lambda_A \psi),
\label{eqn:Lag}
$
with $\psi^C\equiv C\bar{\psi}^T$
and $C = i\gamma_2\gamma_0$ being the charge conjugation operator.
The matrices $\tau_2$ and $\lambda_A$ $(A=2,5,7)$ 
are the antisymmetric components of the Pauli and Gell-Mann matrices 
for the flavor $SU(2)_f$ and color $SU(3)_c$, respectively.
The coupling constants, $G_S$ and $G_C$, and the three-momentum cutoff,
$\Lambda$, are taken from Refs. \citen{Hatsuda:1985eb} and 
\citen{Schwarz:1999dj}.

The fluctuations of the diquark (chiral) condensate are described by
the diquark (quark-antiquark) Green function in the
random phase approximation,
$
  {\cal D}_{C,S}(\bm{p},\nu_n)=-[1/2G_{C,S}+
  {\cal Q}_{C,S}(\bm{p},\nu_n)]^{-1},
$
where the subscript $C (S)$ denotes the diquark (quark-antiquark) sector.
$\nu_n=2\pi nT$ is the Matsubara frequency for bosons and
${\cal Q}_{C,S}(\bm{p},\nu_n)$ is the undressed 
quark-antiquark (diquark) polarization function at one-loop.
To evaluate strengths of the fluctuations, we employ the spectral function,
$\rho$, and the dynamic structure factor, $S$, given by
$
  \rho_{C,S}(\bm{p},\omega) = 
  -(1/\pi){\rm Im}{\cal D}_{C,S}(\bm{p},\nu_n)|_{i\nu_n=i\omega+i\eta}
$ and
$
  S_{C,S}(\bm{p},\omega) =\rho_{C,S}(\bm{p},\omega)/(1-e^{-\omega/T}),$
respectively.
\begin{figure}[t]
\centering
\includegraphics[width=7.0cm]{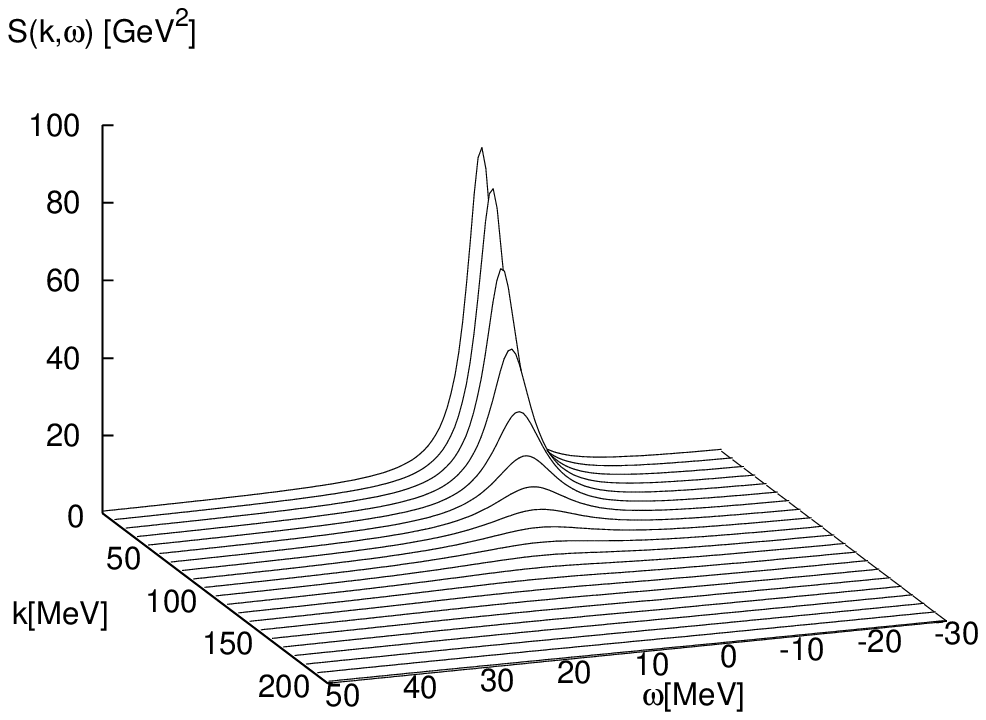}
\includegraphics[width=6.0cm]{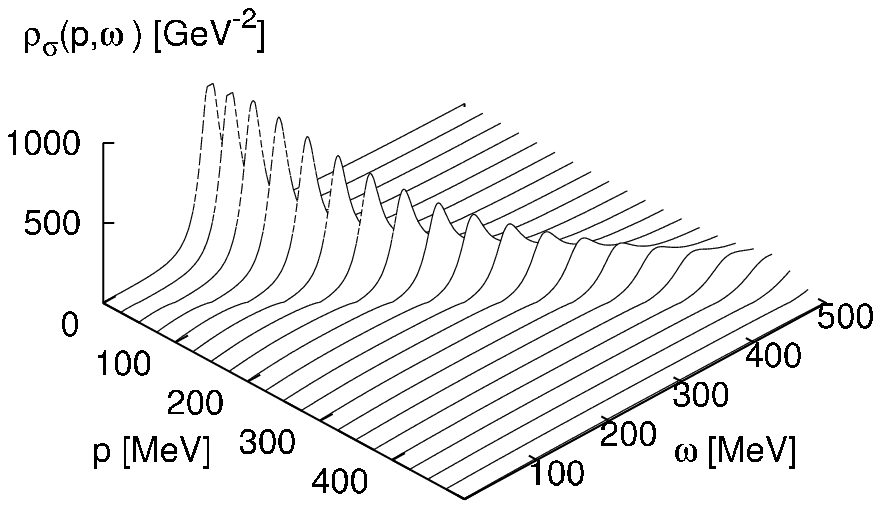}
\caption{The dynamic structure factor $S_C$ for the diquark mode
for $\mu=400$ MeV and $\varepsilon\equiv(T-T_c)/T_c=0.02$ (left), 
and the spectral function $\rho_\sigma\equiv\rho_S$
 for the quark-antiquark mode
for $\mu=0$ and $\varepsilon=0.1$ (right). $k$ and $p$ denote
the momentum.}
\label{fig:imd}
\end{figure}
$S_C(\bm{p},\omega)$ for the diquark mode and 
$\rho_S(\bm{p},\omega)$ for the quark-antiquark mode near
$T_C$ are plotted in Fig. \ref{fig:imd}.
One can see that there appear pronounced peaks which denote
the precursory soft modes.\cite{Hatsuda:1985eb,Kitazawa:2001ft}
The peak positions of these modes are approximately expressed
as
$\omega_C\simeq \bm{p}^2$ for the diquark mode, and 
$\omega_S\simeq \pm \sqrt{m_\sigma^{*}(T)^2+\bm{p}^2}$ for the quark-antiquark
mode.
A $T$-dependent `mass' $m_\sigma^*(T)$ becomes smaller as $T$ approaches $T_c$,
which means the softening at $T_c$.
As will be seen in Sec.\ref{spc}, 
the difference between $\omega_C$ and $\omega_S$
leads to a quite different quark spectrum.

\section{Quark spectrum near the phase transitions} \label{spc}

The effect of the soft modes on the quark spectrum is incorporated in 
the quark self-energy in the non-selfconsistent way,
$  \tilde{\Sigma}_{C,S}(\bm{p},\omega_n) =
  -4T\sum_m \int \frac{d^3q}{(2\pi)^3} {\cal D}_{C,S}(\bm{p}-\bm{q},
  \omega_n-\omega_m){\cal G}_0(\bm{q},\omega_m),
$
where ${\cal G}_0(\bm{q},\omega_m)$ is the free quark propagator with
$\omega_n=(2n+1)\pi T$ being the Matsubara frequency for fermions.
The quasi-quark and quasi-antiquark spectral functions,
$\rho_\pm(\bm{p},\omega)$, are obtained from the retarded self-energies,
$\Sigma^R_\pm(\bm{p},\omega)=(1/2){\rm Tr}[\Sigma^R\gamma^0 \Lambda_\pm]$,
respectively, i.e. $\rho_\pm=-(1/\pi){\rm Im}[\omega+\mu\mp|\bm{p}|-
\Sigma^R_\pm]^{-1}$, with the analytic continuation 
$\Sigma^R(\bm{p},\omega)=
\tilde{\Sigma}(\bm{p},\omega_n)|_{i\omega_n=\omega+i\eta}$ and the projection
operators
$\Lambda_\pm=(1\pm \gamma^0 \bm{\gamma}\cdot\bm{p}/|\bm{p}|)/2$.
In the following,
we show the quark spectral function obtained from the self-energy 
$\tilde{\Sigma}_C
(\tilde{\Sigma}_S)$ for the CS (chiral) transition.

\begin{figure}[t]
\centering
\includegraphics[width=6.0cm]{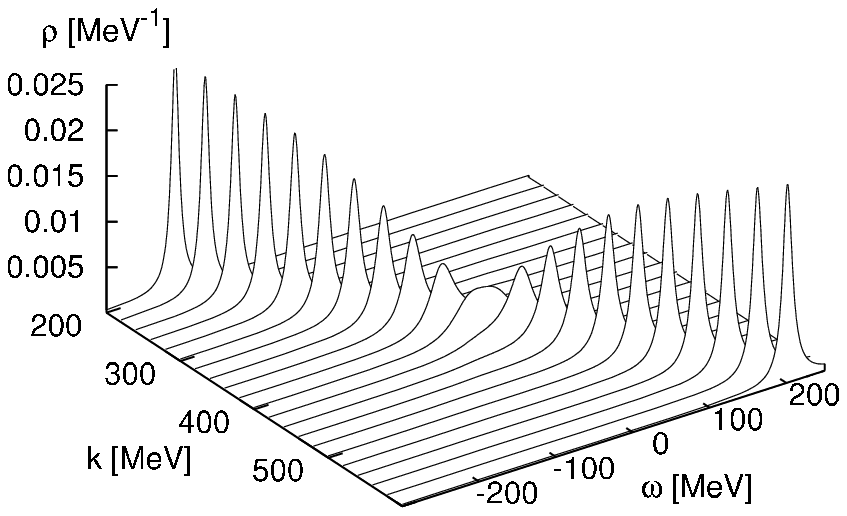}
\includegraphics[width=6.0cm]{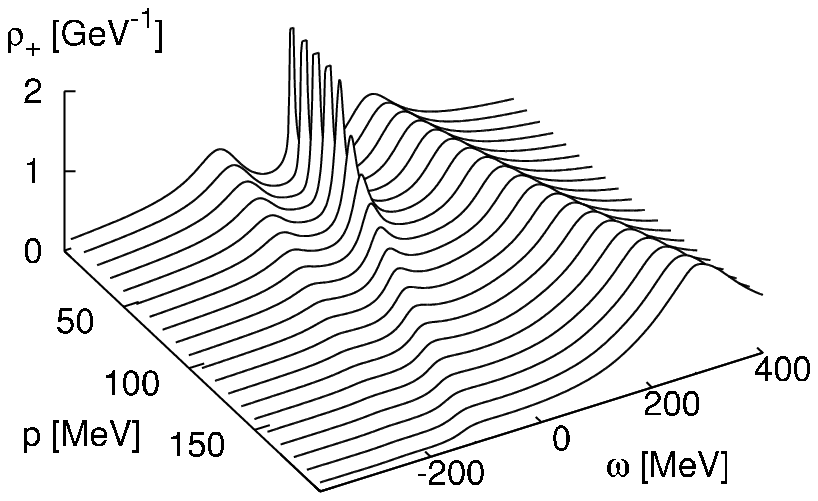}
\caption{The quasi-quark spectral function $\rho_+$ above the CS transition
for $\mu=400$ MeV and $\varepsilon=0.01$ (left), 
and $\rho_+$ above the chiral transition
for $\mu=0$ and $\varepsilon=0.1$ (right). $k$ and $p$ denote
the momentum.}
\label{fig:rho}
\end{figure}
\begin{figure}[t]
\centering
\includegraphics[width=7.0cm]{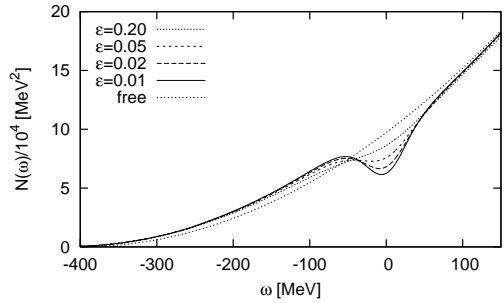}
\caption{The density of states of quarks above the CS transition
 for $\mu=400$ MeV.
The thin dotted curve represents that of free quarks.}
\label{fig:dos}
\end{figure}

\subsection{Color superconducting transition}

The quasi-quark spectral function $\rho_+$ for $\mu=400$ MeV and
$\varepsilon=0.01$ is plotted in the left panel of Fig. \ref{fig:rho}.
We see that the peak has a clear depression around the Fermi energy,
$\omega=0$.\cite{Kitazawa:2003cs}
Owing to the softening of the soft mode near $T_c$, a quark near the Fermi
energy is scattered by the soft mode and create a hole, while a hole can 
create a quark by absorbing the soft mode.
Then, the incident quark and a quark near the Fermi surface make a
{\it resonant scattering} to form the soft mode and vice versa.
This resonant processes induce a virtual mixing between quarks and holes,
which leads to the level repulsion of the energy spectrum near the Fermi 
energy, making the gap-like structure as shown in the left panel of 
Fig. \ref{fig:rho}.
This behavior, which is quite different from the conventional Fermi liquid,
is due to the strong coupling nature between quarks.
In fact, we see that the depression is more significant as the diquark
coupling $G_C$ becomes larger.\cite{Kitazawa:2005pp}

This depression leads to a depression in the density of states (DOS) of quarks
around the Fermi energy, as shown in Fig. \ref{fig:dos}.
Thus, we see a gap-like structure in DOS even above $T_c$, which we call the
pseudogap.\cite{Kitazawa:2003cs}
The non-Fermi liquid behavior is essential for the formation of the pseudogap,
which is analogous to high-$T_c$ superconductors, although the origin of the
pseudogap of the latter is not known precisely.

\subsection{Chiral transition}

The quasi-quark spectral function $\rho_+$ for $\mu=0$ MeV and
$\varepsilon=0.1$ is plotted in the right panel of Fig. \ref{fig:rho}.
We see a clear three-peak structure at low momentum, which
exists even at $\varepsilon=0.2$.\cite{Kitazawa:2005mp}
Although not shown in the figure, the quasi-antiquark spectrum, $\rho_-$,
has also a three-peak structure for a relation,
$\rho_-(\bm{p},\omega)=\rho_+(\bm{p},-\omega)$.
The mechanism of the appearance of the three-peak structure in
$\rho_+$ is as
follows:
The imaginary part of $\Sigma^R_+(\bm{0},\omega)$ has two peaks
at nonzero values of $\omega$, which means that there exist two
large damping modes of the quasi-quark there.
From a kinematical consideration, we see that one is a collision
of a thermally excited antiquark and the quasi-quark creating the soft
mode, and the other is a collision of the quasi-quark and the soft mode
creating an on-shell quark.
Both the processes are interpreted as a Landau damping of the quasi-quark.
The point is that the quasi-quark is a mixed state between quarks and
`antiquark-holes' which are annihilation of thermally excited antiquarks
and have the positive quark number.
Then, these damping modes cause a mixing between quarks and antiquark-holes.
This mixing mechanism can be described in terms of the
resonant scattering as in the case of the color superconductivity,
although a crucial difference arises owing to the different nature of the
soft modes.
We can show that a coupling with the soft mode with 
{\it a nonzero mass $m_\sigma^*(T)$} is
essential for the appearance of the three-peak structure in the quark spectrum.

In fact, we have investigated the quark spectrum in Yukawa models with a 
massive scalar (pseudo-scalar) and vector (axial-vector) boson of a nonzero 
mass $m$, and find that there appears a three-peak structure
in the quark spectral function with a collective nature when temperature is
compatible with $m$.\cite{Kitazawa:2007}
Because the employed Yukawa models are rather generic, the findings
may represent a universal phenomenon for fermions coupled with a massive
bosonic excitation with a vanishing or small width.

\section{Conclusions}

We have investigated
the quark spectrum near the CS and chiral transitions
taking into account the fluctuating soft modes.
We have shown that the quark spectrum shows a non-Fermi liquid behavior
leading to the pseudogap in the density of states above the CS transition,
and a clear three-peak structure above the chiral transition.
The gap-like structure in the spectral function near both the transitions
can be uderstood in terms of the resonant scattering of each soft mode.


%


\end{document}